\documentclass[10pt,letterpaper]{article}
\usepackage[top=0.85in,left=1in,footskip=0.75in,marginparwidth=2in]{geometry}

\usepackage[utf8]{inputenc}

\usepackage{cite}
\usepackage{amsmath}
\usepackage{gensymb}
\usepackage{xfrac}
\usepackage{soul}
\usepackage{xcolor}

\usepackage{nameref}
\usepackage[hidelinks]{hyperref}

\usepackage{microtype}
\DisableLigatures[f]{encoding = *, family = * }

\usepackage{changepage}

\usepackage[aboveskip=1pt,labelfont=bf,labelsep=period,singlelinecheck=off]{caption}

\makeatletter
\renewcommand{\@biblabel}[1]{\quad#1.}
\makeatother

\usepackage{lastpage,fancyhdr,graphicx}
\usepackage{epstopdf}
\fancyheadoffset[L]{2.25in}
\fancyfootoffset[L]{2.25in}

\usepackage{color}

\definecolor{Gray}{gray}{.25}

\usepackage{graphicx}

\usepackage{sidecap}

\usepackage{wrapfig}
\usepackage[pscoord]{eso-pic}
\usepackage[fulladjust]{marginnote}
\reversemarginpar

\begin{document}
\vspace*{0.35in}

% title goes here:
\begin{center}
{\Large
\textbf\newline{Paper-based printed CPW-fed antenna for Wi-Fi applications}
}
\newline
% authors go here:
\\
Nitheesh M. Nair\textsuperscript{1,2},
Debdutta Ray\textsuperscript{2}
Parasuraman Swaminathan\textsuperscript{1,3*}

\bigskip
\textsuperscript{1}Electronic Materials and Thin Films Lab,
Dept. of Metallurgical and Materials Engineering, \\
Indian Institute of Technology, Madras, Chennai, India \\
\textsuperscript{2}Organic Electronics Group,
Dept. of Electrical Engineering, \\
Indian Institute of Technology, Madras, Chennai, India \\
\textsuperscript{3}Ceramics Technologies Group - Center of Excellence in Materials and \\Manufacturing for Futuristic Mobility,
Dept. of Metallurgical and Materials Engineering, \\
Indian Institute of Technology, Madras, Chennai, India
\\
%\bf{2} Affiliation B

\bigskip
*Email: swamnthn@iitm.ac.in

\end{center}

\section*{Abstract}
A paper-based co-planar waveguide (CPW) fed monopole antenna for Wi-Fi applications is proposed. The antenna is fabricated by printing a commercial silver nanoparticle (Ag NP) based ink on photo paper substrate. The antenna is designed as a single layer for the ease of fabrication, and it is designed to radiate at two frequencies, 2.4 and 5.8 GHz, which are suitable for Wi-Fi applications. The printed  film exhibits good electrical conductivity, with a low sheet resistance of 114 m$\Omega/$sq comparable with commonly used conductive metal lines. The fabricated antenna demonstrates good radiative properties at both flat and bent conditions. This confirms the good flexible properties of the antenna making it compatible with mounting on curved surfaces. The performance of the fabricated antenna is also compared with a commercial rigid antenna by interfacing with a USB dongle. The printed antenna demonstrates better performance with respect to signal strength at specific distances when compared with the commercial antenna. This work demonstrates that rigid and long commercial antennas can be replaced with paper-based flexible and cheap antennas and incorporated with wearable technologies. Additionally, replacing Ag NPs with nanowires provides transparency without compromising on the electrical properties. 

\bigskip

\noindent \textbf{Keywords:} Printed electronics; Antenna; Paper electronics; Wi-Fi; Co-planar waveguide

\bigskip
% now start line numbers
%\linenumbers
\newpage
% the * after section prevents numbering
\section {Introduction}
Paper is one of the most widely used, inexpensive, and commonly available materials\cite{Prebianto2018paper}. It has recently been used extensively as a flexible substrate for fabricating electronic devices, more popular as ``paper-based electronics''\cite{Valentini2013flexible, Liu2021flexible, Zhang2021integrating, Li2014direct}. It possesses excellent biodegradable properties. A variety of devices can be implemented on paper substrates\cite{Yang2007rfid, Cook2012inkjet, Kim2013inkjet} and among them wireless wearable sensors have great potential in real-time health monitoring, sports and fitness, military applications, and entertainment\cite{Chen2019wireless}. An RF front end is an essential primary component in any communication system, including passive components such as antennas, transmission lines, and impedance matching networks and active components such as oscillators and amplifiers\cite{Huang2015highly}. Among them, the antenna can convert electrical energy to electromagnetic signals and vice versa. For wearable applications the antenna needs to be flexible and at the same time efficiently radiate the desired frequency\cite{Nair2020}. Conventionally, an antenna is fabricated by copper etching or by photolithography techniques, which are not compatible with paper electronics and will increase the overall manufacturing costs\cite{Li2014direct}.

Printed electronics has gained wide attention due to its low cost, ease of fabrication, and compatibility with substrates with low thermal budgets, such as paper, cellulose, and polyethylene terephthalate (PET)\cite{Radivojevic2007optimised, Nair2021}. Among the various printing techniques available, inkjet printing and direct writing offer significant advantages in terms of their versatility\cite{Sharma2017top}. Direct writing is an additive manufacturing technique where the material is selectively deposited (typically by extrusion process) in the required areas, by a drop-on-demand process\cite{Balani2021processes}. This helps in minimizing material wastage and allows fabrication in ambient conditions. For printing, a stable dispersion of nanostructures, typically nanoparticles (NPs), in a suitable solvent is required with predefined rheological properties\cite{Sharma2017top, Suganthi2018formulation}. Noble metals, such as Ag, Cu, and Au, are highly conducting materials and suitable to make conductive printable NP-based inks. Among these, Ag has been widely used because of its high bulk electrical conductivity, better oxidation stability compared to Cu, and cost-effectiveness compared to Au NPs\cite{Bourassa2019, Karim2019inkjetprinted}. Along with NPs, copper, silver, and gold nanowires are also used, primarily for printing applications\cite{Nair2019, Nair2020, Nair2021FPE, Nam2016, Chirea2011, Ganapathi2019anodic, Kumaresh2021templateassisted}. The printing technology can be employed to fabricate antennas patterns on paper using metallic NP inks\cite{Pakkathillam2021planar}.

Wi-Fi and Bluetooth are two popular communication standards, commonly available in devices such as cellphones and laptops, which are widely used for wearable applications. The standard Wi-Fi protocols IEEE 802.11 a/b/g/ac/n, and Bluetooth protocol IEEE 802.15.1 uses 2.4GHz and 5 GHz (5.2 and 5.8 GHz) frequencies as their centre frequencies with a narrow bandwidth\cite{Alam2016wearable}. The electronic devices for all Wi-Fi and Bluetooth protocols need a dual-band antenna that can radiate at 2.4 GHz and 5 GHz of frequencies\cite{Hassan2017inkjet}. In this work, we describe the design and fabrication of a co-planar waveguide (CPW) fed monopole antenna that satisfies the 2.4 GHz and 5.8 GHz application requirements. The antenna is designed using CST studio suite and fabricated by direct writer printing using a commercial Ag NP-based ink. The performance of the antenna is evaluated under bent conditions and also compared with a commercial rigid antenna. The work demonstrates that printed paper-based flexible antennas can serve as low-cost alternatives to commercial long and rigid antennas and can be easily integrated for wearable applications.

\section {Materials and methods}
\subsection{Antenna Design}
\begin{figure}[h]
    \centering
    \includegraphics[width=14cm]{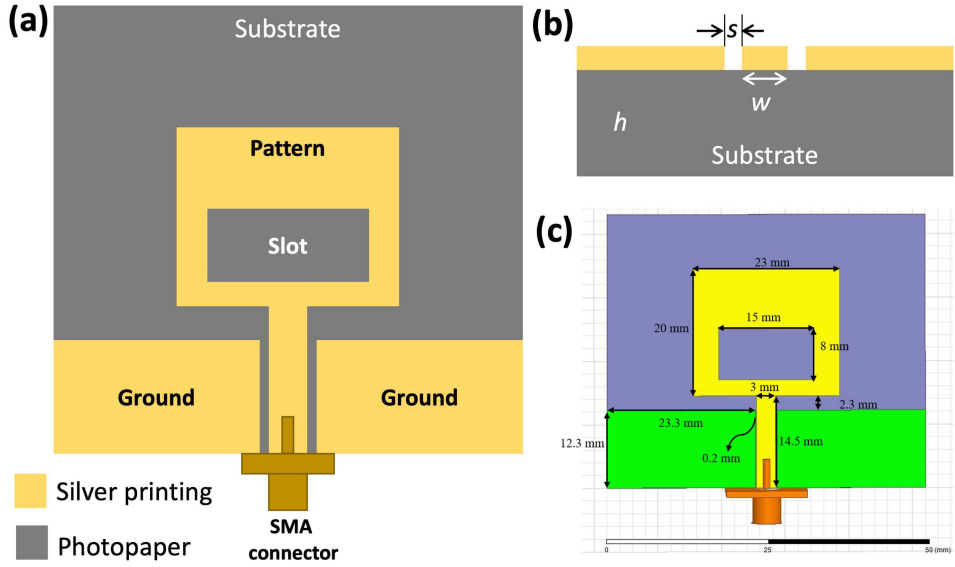}
    \caption{a) Schematic of the CPW fed antenna design. b) Structure of the co-planar waveguide design. c) Dimensions of the optimised antenna.}
    \label{fig:Fig1}
\end{figure}

Figure \ref{fig:Fig1}(a) shows the antenna pattern designed using CST. The antenna structure consists of a radiating rectangular monopole pattern with a slot inside, fed using a CPW. The main advantage of designing a CPW feeding structure is that the radiating element and the ground plane can be fabricated on the same plane of the substrate, which helps for low-cost, fast, printed fabrication\cite{Hassan2017inkjet, Jo2014cpw, Danideh2013cpw}. Also, this structure has low radiation loss, high impedance matching, and wide bandwidth characteristics. A Sub miniature version A (SMA) connector with 50 $\Omega$ impedance was used to interface with the antenna.

A CPW is a planar strip transmission line with various conducting structure arrays configured in a single geometric plane, used to transmit high-frequency signals. It consists of a conducting metallic strip deposited on a dielectric material and two ground planes adjacent and parallel to the strip on both sides in the same plane with a small slit in between. The characteristic impedance ($Z_0$) of the CPW transmission line is given by\cite{Edwards2016foundations}
\begin{equation}\label{eq:Eqn1}
    Z_0 \: =\: \dfrac{30 \pi}{\sqrt{\epsilon_{eff}}}\,\dfrac{K(k')}{K(k)}
\end{equation}
where $k$ and $k'$ are given by
\begin{equation}
\begin{split}
    k\: = \: \dfrac{s}{w+s}\\
    k' \: =\: \sqrt{(1\,-\,k^2)}
\end{split}
\end{equation}
Here, $w$ is the width of the centre strip and $s$ is the spacing between the centre strip and the ground plane. $K(k)$ and $K(k')$ represent the elliptical integral
\begin{equation}
    K(k)\: = \: \int_0^{\pi/2}\dfrac{d\phi}{\sqrt{1\,-\,k^2\sin^2\phi}}
\end{equation}
The dimensions of the CPW structure ($w$ and $s$) (see figure \ref{fig:Fig1}(b)) were adjusted to match its impedance with that of the SMA connector, i.e., 50 $\Omega$. Based on the dielectric constant of the substrate, the width of the middle strip and the slit gap between strip and ground were adjusted to tune the impedance. If there is any mismatch, a finite fraction of the radiating power will be reflected back to the feeding circuit, resulting in a reduced efficiency\cite{Chaimool2012cpw}. 

To achieve the dual-band characteristics, the dimensions of the slot and the rectangular pattern are adjusted. The dimensions of the pattern are related to the guide wavelength ($\lambda_g$) given by\cite{Edwards2016foundations}
\begin{equation} \label{eq:Eqn4}
    \lambda_g \: =\: \dfrac{c}{f\sqrt{\epsilon_{eff}}}
\end{equation}
where $c$ is the speed of light, $f$ is the signal frequency for which the antenna is designed, and $\epsilon_{eff}$ is the effective dielectric constant given by
\begin{equation}
    \epsilon_{eff}\: =\: \dfrac{\epsilon_{sub}\,+\,1}{2}
\end{equation}
For photopaper, $\epsilon_{sub}\,=\,3.2$, the substrate thickness is 0.22 mm and loss tangent is 0.05\cite{Abutarboush2015inkjet}. For these values, $\epsilon_{eff}\,=\,2.1$ and $\lambda_g$ for 2.4 GHz is 86.25 mm. So the $\lambda_g/4$ will be 22 mm, which is taken as the initial dimension for the rectangular pattern length. Then a slot is introduced to achieve the dual band. The sizes of the pattern as well as the dimensions and position of slot were adjusted to get an optimised antenna pattern that radiates at 2.4 GHz and 5.8 GHz frequencies. The optimised dimensions are given in figure \ref{fig:Fig1}(c).

\subsection{Experimental Details}

The antenna pattern was simulated using the CST Studio suite and Ansoft HFSS. Then the antenna was printed using a custom-built direct writer manufactured by Tvasta Manufacturing Solutions Pvt. Ltd. The printer consists of a syringe based head for ink extrusion and a stepper motor for motion control\cite{Suganthi2018formulation}. Ag NP ink, purchased from Methode Electronics, was used to print on Kodak ultra-premium photo paper as the substrate. Printing was done at room temperature and after printing, the pattern was annealed at 100 \degree C for 1 h in a hot air oven. Post annealing, the SMA connector was cold-soldered to the pattern using silver epoxy from MG chemicals and allowed to dry at ambient for 24 h.

Transmission electron microscopy (TEM) images of the NPs were captured using a Philips CM 12 microscope, operating at 120 kV. The contact angle measurement was performed using an Apex Acam-D3 contact angle meter. The morphology of the film was observed using scanning electron microscopy (SEM), FEI Quanta 400, operating at 20 kV and a Metallux 3 Leitz Wetzlar microscope. The sheet resistance was measured using Jandel RM3000 four-probe instrument and thickness and surface profile were mapped using Nanomap 1000 WLI optical profilometer from AEP Technology. The reflection coefficient of the antenna under test was measured using Agilent E5071C Vector Network Analyser (VNA). 

\section {Results and Discussion}
\subsection{Simulation studies on the effect of antenna geometry}
\begin{figure}
    \centering
    \includegraphics[width=14cm]{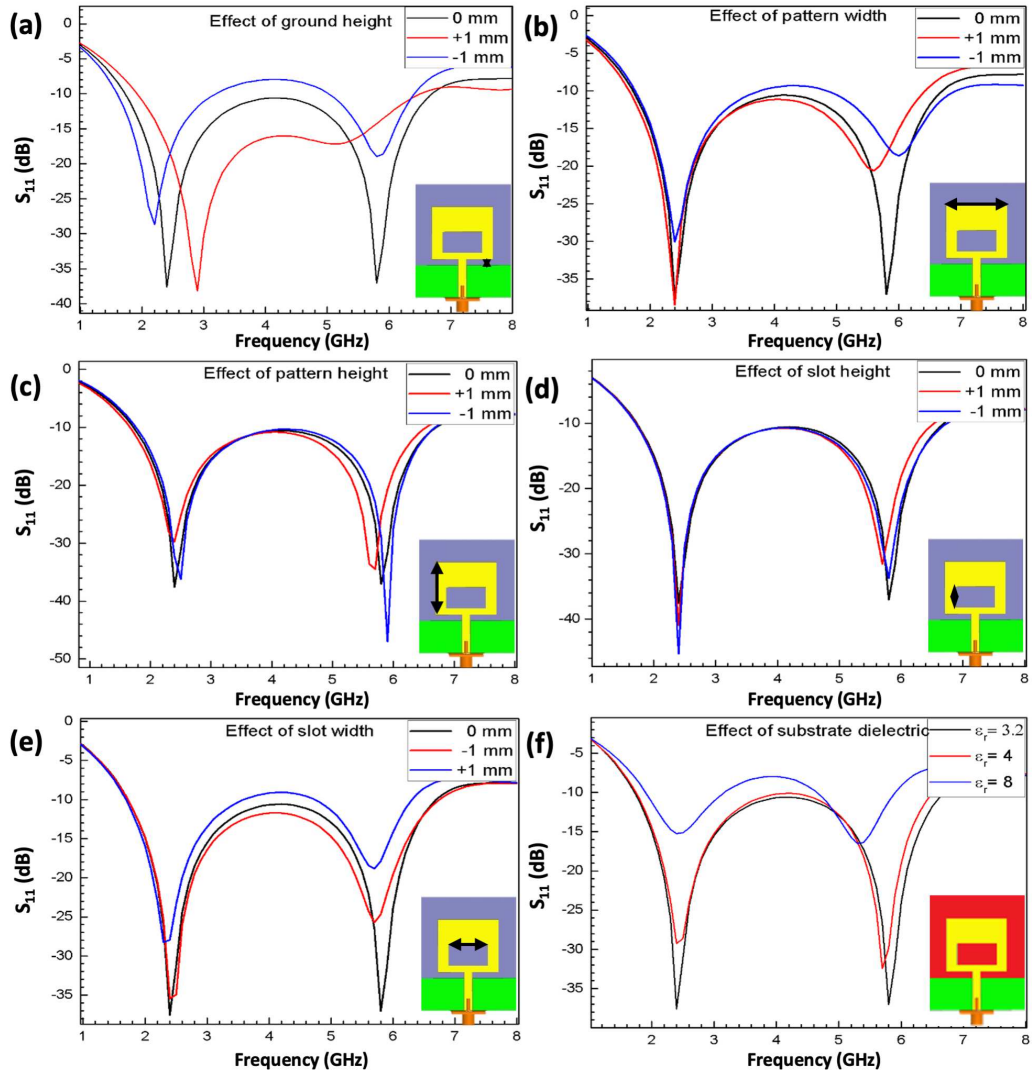}
    \caption{Effect on the antenna parameters on the reflection coefficient. The variation of the reflection coefficient with a) height of the ground pattern, b) width of the pattern, c) pattern height, d) slot height, e) slot width, and f) substrate dielectric properties. The simulations were performed using CST studio suite.}
    \label{fig:Fig2}
\end{figure}

To optimise the antenna pattern geometry and to investigate the performance of the proposed design, the antenna pattern was simulated using CST studio suite. Figure \ref{fig:Fig2} shows the simulated data that models the effect of different antenna dimensions on the reflection coefficient, which measures the amount of signal reflected from the antenna, i.e., the unradiated part of the signal. Figure \ref{fig:Fig2}(a) shows the relation between the rectangular pattern and the ground height. It can be used to control the lower centre frequency and impedance matching at the higher frequency band. Figure \ref{fig:Fig2}(b) confirms that the width of the rectangular pattern will primarily affect the higher frequency value. Figure \ref{fig:Fig2}(c) shows the effect of return loss with a change in pattern height, keeping the pattern width and slot dimensions constant. Both higher and lower frequencies were reduced with a reduction in pattern height. It also has a role in setting the return loss in the lower band. Figures \ref{fig:Fig2}(d) and (e) show that the slot dimensions mainly control the return losses and also help to fine-tune both frequencies. Equation \ref{eq:Eqn4} shows that with a higher dielectric constant of the substrate, the radiating frequency will reduce, the factor is again confirmed using the simulation studies (figure \ref{fig:Fig2}(f)), as the centre frequency is reduced with dielectric constant with the same antenna dimensions. The return loss is also increased, due to a change in CPW impedance with the dielectric constant (equation \ref{eq:Eqn1}), as CPW dimensions remain the same.   

\begin{figure}
    \centering
    \includegraphics[width=14cm]{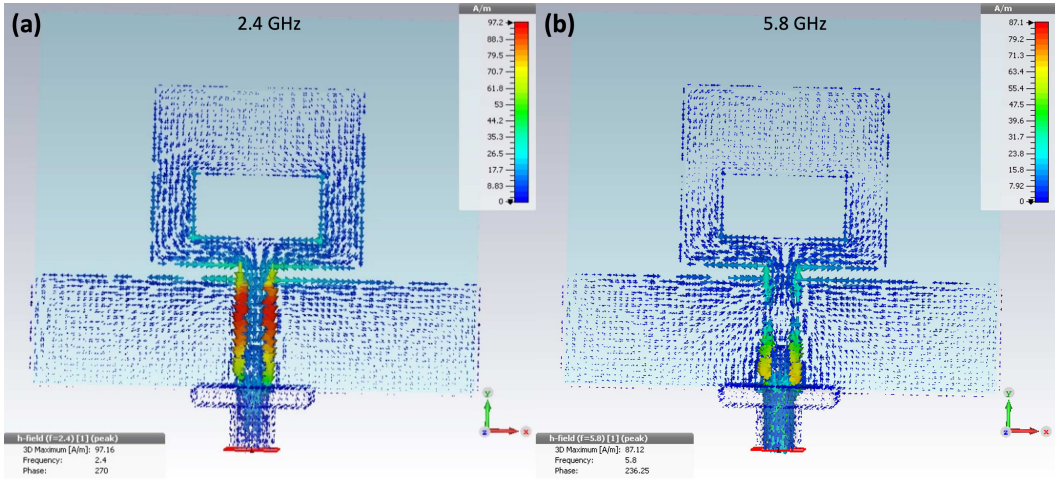}
    \caption{Simulated current distribution along the optimised antenna pattern at a) 2.4 GHz and b) 5.8 GHz.}
    \label{fig:Fig3}
\end{figure}

\begin{figure}
    \centering
    \includegraphics[width=14cm]{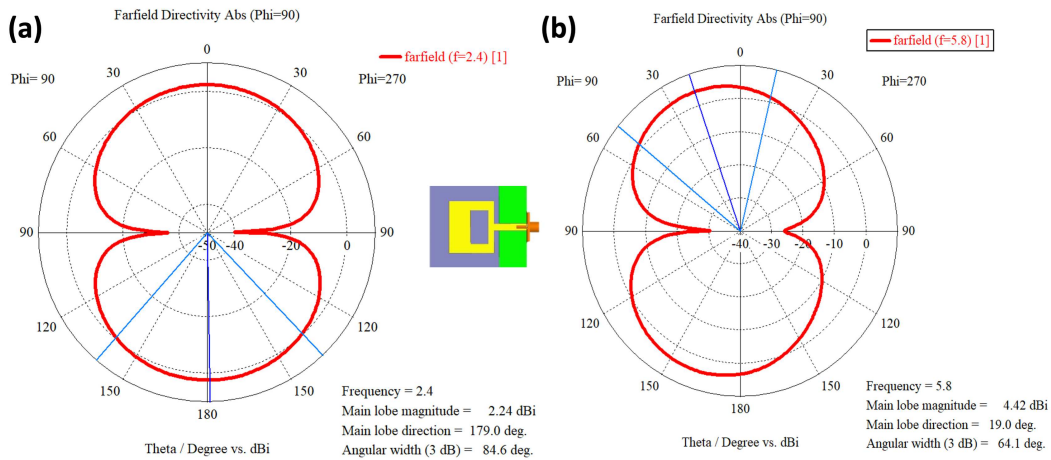}
    \caption{The simulated far-field radiation pattern of the antenna at a) 2.4 GHz and b) 5.8 GHz.}
    \label{fig:Fig4}
\end{figure}

The simulated current distributions at various frequencies in the optimised antenna pattern are shown in figure \ref{fig:Fig3}. Based on the resonant frequency, the current distribution varies significantly. At 2.4 GHz, the current is focused on the vertical stripe of the antenna pattern (figure \ref{fig:Fig3}(a)). Similarly, at 5.8 GHz, the current is concentrated in the horizontal region at the bottom of the rectangular pattern (figure \ref{fig:Fig3}(b)). So, by adjusting these two dimensions, we can adjust the current distribution at different frequencies and thereby change the resonant radiation frequencies of the pattern. Figure \ref{fig:Fig4} plots the simulated far-field radiation patterns, which shows how the antenna radiates the electromagnetic signals to the outside environment. The electric field patterns exhibit some lobes with peak gains of 2.24 dBi and 4.42 dBi at 2.4 GHz and 5.8 GHz respectively.  

\subsection{Printed antenna}
\begin{figure}
    \centering
    \includegraphics[width=14cm]{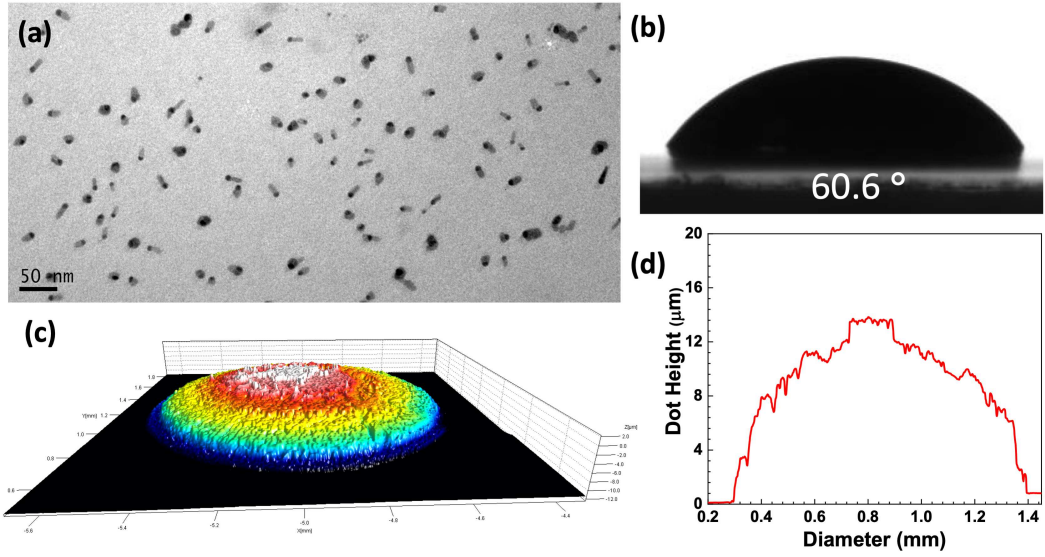}
    \caption{a) TEM of the Ag NPs dispersed on the ink. b) Contact angle made by the ink on the photo paper substrate. c) The surface profile of the dried drop pattern of the ink and d) its corresponding height profile.}
    \label{fig:Fig5}
\end{figure}

After simulation, the optimised antenna pattern was fabricated by direct writer printing technique using Ag NP ink (figure \ref{fig:Fig5}(a) shows the TEM image of the Ag NPs) on photo paper. For ensuring proper jettability of NP based inks, there are certain physical requirements primarily set by parameters such as the viscosity, surface tension, and the nozzle diameter of the printer\cite{Sharma2017top}. The nozzle diameter is fixed for a given printer, and other parameters can be adjusted by tuning the rheological properties of the ink during the formulation. For a drop-on-demand printer, the viscosity must lie between 2 to 20 mPa. s\cite{Suganthi2018formulation, Sharma2017top}. A lower value of viscosity may lead to the dripping of ink from the nozzle head and satellite drop formation, while higher viscosity values prevent free ink flow and will require a larger extrusion force. An adequate surface tension, between 35 to 70 mN/m, ensures that ink droplets are formed at the nozzle without dripping and spread across the substrate with uniformity\cite{Derby2010inkjet}. To define printability of an ink, a dimensionless number, Ohnesorge number ($Oh$), is used, which is written as
\begin{equation}\label{eq:Eq6}
    Oh \: =\: \dfrac{\eta}{\sqrt{\sigma \rho l}}
\end{equation}
$\eta$ is the viscosity, $\sigma$ is the surface tension, and $\rho$ is the mass density of the fluid. The nozzle diameter is given by $l$. $Oh$ must lie between 0.1 and 0.5 for good printability\cite{Nair2019, Nair2020, Suganthi2018formulation}. The Ag NP ink used for the antenna fabrication in this work has a viscosity of 3.2 mPa. s, surface tension of 44.7 mN/m, and mass density of 1.2 g/ml.  Taking the appropriate value of nozzle diameter (200 $\mu$m) and using equation \ref{eq:Eq6} the $Oh$ value was found to be approximately 0.14, satisfying the jettability criteria. 

During printing, the ink must properly wet the substrate surface, and the wettability properties can be understood using contact angle studies. A contact angle of less than 90\degree, i.e, a hydrophilic surface, is desired, but at the same time, a too low a value of contact angle results in spreading of the ink and loss of pattern fidelity. The Ag NP ink forms a contact angle of 60.6\degree$\,$ with photo paper (figure \ref{fig:Fig5}(b)), ensuring good pattern fidelity on the paper substrate. Self-assembly of the NPs while drying is also very important, as the particles tend to agglomerate at the boundaries of the pattern due to the outward solvent flow while drying, termed the coffee ring effect\cite{Suganthi2018formulation}. Figures \ref{fig:Fig5}(c) and (d) shows the surface profile of a single droplet. A uniform 3D hemispherical assembly of NPs was observed, confirming the absence of the coffee ring effect.

\begin{figure}
    \centering
    \includegraphics[width=14cm]{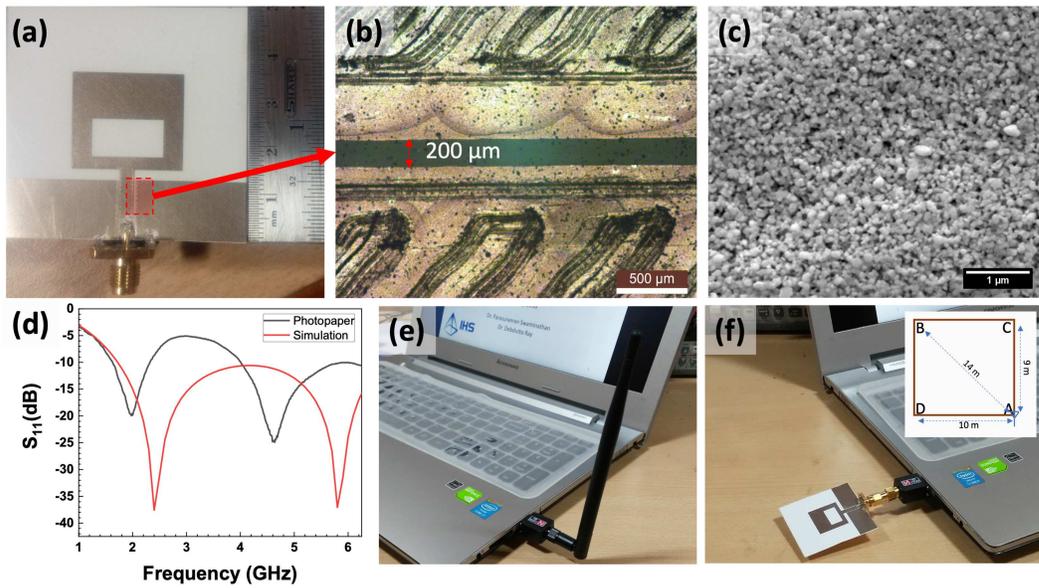}
    \caption{a) Image of the printed antenna on photopaper. b) Optical micrograph of the gap between the centre stripe and ground. A gap of 200 $\mu$m has been achieved, confirming the precision in printing. c) SEM of the printed Ag NP based ink on photo paper after annealing. d) S$_{11}$ characteristics of the simulated and fabricated paper antenna. e) Commercial and f) printed antenna connected to the USB dongle to set up the Wi-Fi hotspot for comparison. Inset showing the position at which the antenna strength was measured. The antenna was installed at A, and the signal strength was measured using an android app, Network analyser lite, at positions A to D. The signal strength data is shown in Table \ref{tab:Table1}.}
    \label{fig:Fig6}
\end{figure}

The printed pattern shows a sheet resistance of 114 m$\Omega$/sq., which is suitable for most printed electronics applications. Figure \ref{fig:Fig6}(a) shows the image of the printed Ag NP antenna on photo paper. The desired spacing of 200 $\mu$m between the centre stripe and the ground was successfully achieved, which is shown in the optical micrograph in figure \ref{fig:Fig6}(b), confirming the precision of the fabrication. The porous nature of the printed film can be seen in the SEM image in figure \ref{fig:Fig6}(c) mainly due to the porous nature of the photopaper and due to the evaporation of the solvents from the ink during film formation.

\begin{figure}
    \centering
    \includegraphics[width=14cm]{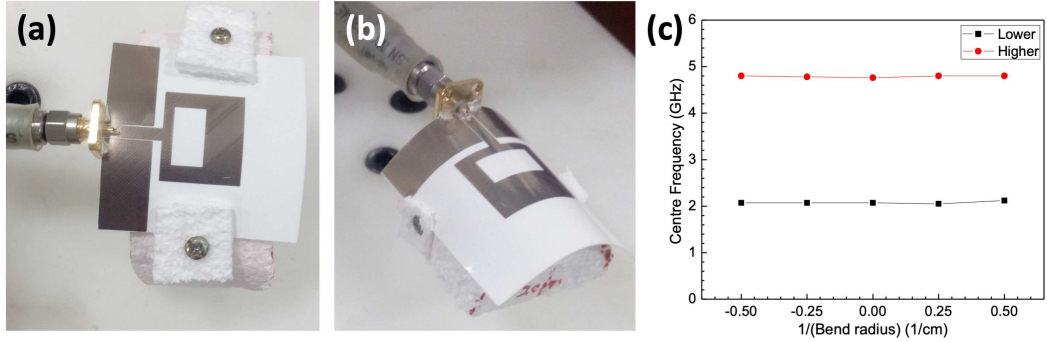}
    \caption{The photograph of the antenna in bend conditions a) top view and b) parametric view. c) The variation in the centre frequency with bending. The variation is negligible.}
    \label{fig:Fig7}
\end{figure}

Figure \ref{fig:Fig6}(d) shows the measured reflection coefficient of the fabricated antenna. The dual-band frequency was exhibited clearly and matched well with the simulated data. The paper antenna exhibited a reflection coefficient of $-19.7$ and $-24.7$ dB at 2.4 and 5.8 GHz frequencies, respectively, confirming good impedance matching. A red shift (shift to lower values) is observed in the centre frequency, which could be due to the change in the dielectric constant of the substrate, especially at the operating (higher) frequencies. The surface roughness of the photo paper is high, with holes and pits present, which will add some surface roughness and surface inductance terms that can also contribute to the shift\cite{Sim2017rf}. The connector losses, including the losses in the SMA and the presence of Ag epoxy, are also significant, which was not considered in the simulation\cite{Chakraborty2019templated}. The high bandwidth of 0.71 and 1.96 GHz at the lower and higher frequencies, respectively, guarantee good reception and transmission of Wi-Fi standard frequencies. The radiation behaviour of the fabricated paper-based antenna was compared with a commercial antenna (figure \ref{fig:Fig6}(e)), manufactured by ShopAIS, by connecting with a USB Wi-Fi dongle. The fabricated antenna (figure \ref{fig:Fig6}(f)) was observed to receive the Wi-Fi signals, similar to the commercial antenna. To verify the transmitting properties, a Wi-Fi hotspot was created using both the antennas and measuring the signal strength, tabulated in Table \ref{tab:Table1}, at various distances using an android based application, `Network analyser lite'. Better signal strengths (lower negative values) were observed in most of the positions confirm that the paper antenna performs better compared to the commercial antenna. The antenna performance was also evaluated under inward and outward bent conditions, by attaching to fabricated curvatures of 2 and 4 cm in diameter to exert tensile and compressive stress, as shown in figures \ref{fig:Fig7}(a) and (b). The centre frequencies were observed to be the same after multiple bend tests, as seen in figure \ref{fig:Fig7}(c), thus confirming the flexibility of the antenna. Thus, the long rigid commercial antennas can be replaced with our flexible paper based antennas in the future. Similarly, replacing silver NPs with nanowire-based inks can also make these printed antennas transparent\cite{Nair2020} which can further improve their applicability. 

\begin{table}[h]
    \centering
    \caption{The signal strength from the commercial and printed antennas was measured at spatial positions A-D (see inset of figure \ref{fig:Fig6}(f)).  \\}
    \begin{tabular}{|c|c|c|c|c|}
    \hline
    \textbf{Antenna (dBm)}&\textbf{A (0 m)} & \textbf{B (14 m)} & \textbf{C (9 m)} & \textbf{D (10 m)}  \\
    \hline
\textbf{Commercial} & $-32$ & $-49$ & $-45$ & $-50$\\
\textbf{Paper antenna} & $-29$ & $-44$ & $-49$ & $-41$\\
         \hline
    \end{tabular}
    \label{tab:Table1}
\end{table}

\newpage

\section{Conclusion}

In this study, we fabricated a flexible dual-band CPW fed monopole antenna by printing technique using Ag NP ink on photo paper, designed for Wi-Fi applications.  The RF characteristics of the antenna were simulated and measured and agree well with the simulated values, with a slight red shift mainly dominated by the variation in the photo paper dielectric constant at the frequencies of interest. The printed Ag films exhibited a very low sheet resistance of 114 m$\Omega$/sq. The fabricated antenna was observed to be radiating under both flat and bent conditions and hence suitable for flexible applications. The fabricated antenna has a better performance in comparison with a commercial antenna. Therefore rigid and long commercial antennas can be replaced with paper-based flexible and cheaper antennas in future.

\section*{Acknowledgments}
Support from the Ceramics Technologies Group - Center of Excellence in Materials and Manufacturing for Futuristic Mobility (project number SB/2021/0850/MM/MHRD/008275) is also acknowledged. We would also like to acknowledge the Liquid film coatings lab, Department of Metallurgical and Materials Engineering, IIT Madras for contact angle meter facility (Funded by DST-SERB, Project No: EMR/2016/ 001479, PI: Sreeram K Kalpathy) to measure the contact angle and surface tension. The viscosity measurements were done in the Polymer Engineering and Colloidal Science (PECS) Lab, Department of Chemical Engineering, IIT Madras. The VNA measurements were performed in the Electromagnetic Research Laboratory, Department of Engineering Design, IIT Madras.

%\nolinenumbers

%This is where your bibliography is generated. Make sure that your .bib file is actually called library.bib
\bibliography{library}

\begin{thebibliography}{10}

\bibitem{Prebianto2018paper}
Nanta~Fakih Prebianto and Asrizal~Deri Futra.
\newblock Paper as a substrate for sensor applications: A review.
\newblock In {\em 2018 International Conference on Applied Engineering (ICAE)},
  pages 1--5. IEEE, 2018.

\bibitem{Valentini2013flexible}
Luca Valentini, Marta Cardinali, Mirjana Mladjenovic, Petar Uskokovic, Federico
  Alimenti, Luca Roselli, and Josè Kenny.
\newblock Flexible transistors exploiting p3ht on paper substrates and graphene
  oxide films as gate dielectrics: proof of concept, 2013, 1302.1291.

\bibitem{Liu2021flexible}
Xiangli Liu, Ziheng Ye, Ling Zhang, Pengdong Feng, Jian Shao, Mao Zhong, Zheng
  Chen, Lijie Ci, Peng He, Hongjun Ji, and et~al.
\newblock Highly flexible electromagnetic interference shielding films based on
  ultrathin ni/ag composites on paper substrates.
\newblock {\em Journal of Materials Science}, 56(9):5570–5580, Jan 2021.

\bibitem{Zhang2021integrating}
Wenliang Zhang, Qinghua Zhao, Carmen Munuera, Martin Lee, Eduardo Flores,
  João~E.F. Rodrigues, Jose~R. Ares, Carlos Sanchez, Javier Gainza, Herre~S.J.
  van~der Zant, and et~al.
\newblock Integrating van der waals materials on paper substrates for
  electrical and optical applications.
\newblock {\em Applied Materials Today}, 23:101012, Jun 2021.

\bibitem{Li2014direct}
Ruo-Zhou Li, Anming Hu, Tong Zhang, and Ken~D Oakes.
\newblock Direct writing on paper of foldable capacitive touch pads with silver
  nanowire inks.
\newblock {\em ACS applied materials \& interfaces}, 6(23):21721--21729, 2014.

\bibitem{Yang2007rfid}
Li~Yang, Amin Rida, Rushi Vyas, and Manos~M Tentzeris.
\newblock Rfid tag and rf structures on a paper substrate using inkjet-printing
  technology.
\newblock {\em IEEE transactions on microwave theory and techniques},
  55(12):2894--2901, 2007.

\bibitem{Cook2012inkjet}
Benjamin~S Cook and Atif Shamim.
\newblock Inkjet printing of novel wideband and high gain antennas on low-cost
  paper substrate.
\newblock {\em IEEE Transactions on Antennas and Propagation},
  60(9):4148--4156, 2012.

\bibitem{Kim2013inkjet}
Sangkil Kim, Benjamin Cook, Taoran Le, James Cooper, Hoseon Lee, Vasileios
  Lakafosis, Rushi Vyas, Riccardo Moro, Maurizio Bozzi, Apostolos Georgiadis,
  et~al.
\newblock Inkjet-printed antennas, sensors and circuits on paper substrate.
\newblock {\em IET microwaves, antennas \& propagation}, 7(10):858--868, 2013.

\bibitem{Chen2019wireless}
Ang Chen, Andrew~Joshua Halton, Rachel~Diane Rhoades, Jayden~Charles Booth,
  Xinhao Shi, Xiangli Bu, Ning Wu, and Junseok Chae.
\newblock Wireless wearable ultrasound sensor on a paper substrate to
  characterize respiratory behavior.
\newblock {\em ACS sensors}, 4(4):944--952, 2019.

\bibitem{Huang2015highly}
Xianjun Huang, Ting Leng, Mengjian Zhu, Xiao Zhang, JiaCing Chen, KuoHsin
  Chang, Mohammed Aqeeli, Andre~K Geim, Kostya~S Novoselov, and Zhirun Hu.
\newblock Highly flexible and conductive printed graphene for wireless wearable
  communications applications.
\newblock {\em Scientific reports}, 5(1):1--8, 2015.

\bibitem{Nair2020}
Nitheesh~M Nair, Jayaram~Kizhekke Pakkathillam, Krishna Kumar, Kavitha
  Arunachalam, Debdutta Ray, and Parasuraman Swaminathan.
\newblock Printable silver nanowire and pedot: Pss nanocomposite ink for
  flexible transparent conducting applications.
\newblock {\em ACS Applied Electronic Materials}, 2(4):1000--1010, 2020.

\bibitem{Radivojevic2007optimised}
Z.~Radivojevic, K.~Andersson, K.~Hashizume, M.~Heino, M.~Mantysalo,
  P.~Mansikkamaki, Y.~Matsuba, and N.~Terada.
\newblock Optimised curing of silver ink jet based printed traces, 2007,
  0709.1842.

\bibitem{Nair2021}
Nitheesh~M Nair, Ishani Khanra, Debdutta Ray, and Parasuraman Swaminathan.
\newblock Silver nanowire-based printable electrothermochromic ink for flexible
  touch-display applications.
\newblock {\em ACS Applied Materials \& Interfaces}, 13(29):34550--34560, 2021.

\bibitem{Sharma2017top}
Sonia Sharma, Sumukh~S Pande, and P~Swaminathan.
\newblock Top-down synthesis of zinc oxide based inks for inkjet printing.
\newblock {\em RSC advances}, 7(63):39411--39419, 2017.

\bibitem{Balani2021processes}
Shahriar~Bakrani Balani, Seyed~Hamidreza Ghaffar, Mehdi Chougan, Eujin Pei, and
  Erdem {\c{S}}ahin.
\newblock Processes and materials used for direct writing technologies: A
  review.
\newblock {\em Results in Engineering}, 11:100257, 2021.

\bibitem{Suganthi2018formulation}
KS~Suganthi, K~Harish, Nitheesh~M Nair, and P~Swaminathan.
\newblock Formulation and optimization of a zinc oxide nanoparticle ink for
  printed electronics applications.
\newblock {\em Flexible and Printed Electronics}, 3(1):015001, 2018.

\bibitem{Bourassa2019}
Justin Bourassa, Alex Ramm, James~Q. Feng, and Michael~J. Renn.
\newblock Water vapor-assisted sintering of silver nanoparticle inks for
  printed electronics.
\newblock {\em SN Applied Sciences}, 1(6), May 2019.

\bibitem{Karim2019inkjetprinted}
Nazmul Karim, Shaila Afroj, Sirui Tan, Kostya~S. Novoselov, and Stephen~G.
  Yeates.
\newblock All inkjet-printed graphene-silver composite inks for highly
  conductive wearable e-textiles applications, 2019, 1905.00839.

\bibitem{Nair2019}
Nitheesh~M Nair, Kevin Daniel, Sai~Chandrahaas Vadali, Debdutta Ray, and
  P~Swaminathan.
\newblock Direct writing of silver nanowire-based ink for flexible transparent
  capacitive touch pad.
\newblock {\em Flexible and Printed Electronics}, 4(4):045001, 2019.

\bibitem{Nair2021FPE}
Nitheesh~M Nair, Mohammad~Mahaboob Jahanara, Debdutta Ray, and P~Swaminathan.
\newblock Photoresponse of a printed transparent silver nanowire-zinc oxide
  nanocomposite.
\newblock {\em Flexible and Printed Electronics}, 6(4):045004, 2021.

\bibitem{Nam2016}
Vu~Binh Nam and Daeho Lee.
\newblock Copper nanowires and their applications for flexible, transparent
  conducting films: a review.
\newblock {\em Nanomaterials}, 6(3):47, 2016.

\bibitem{Chirea2011}
Mariana Chirea, Andreia Freitas, Bogdan~S Vasile, Cristina Ghitulica, Carlos~M
  Pereira, and Fernando Silva.
\newblock Gold nanowire networks: synthesis, characterization, and catalytic
  activity.
\newblock {\em Langmuir}, 27(7):3906--3913, 2011.

\bibitem{Ganapathi2019anodic}
Arulkumar Ganapathi, P~Swaminathan, and Lakshman Neelakantan.
\newblock Anodic aluminum oxide template assisted synthesis of copper nanowires
  using a galvanic displacement process for electrochemical denitrification.
\newblock {\em ACS Applied Nano Materials}, 2(9):5981--5988, 2019.

\bibitem{Kumaresh2021templateassisted}
Kumresh~K R, Lakshman Neelakantan, and Parasuraman Swaminathan.
\newblock Template-assisted growth of silver nanowires by electrodeposition,
  2021, 2201.04947.

\bibitem{Pakkathillam2021planar}
Jayaram~Kizhekke Pakkathillam, Nitheesh~M Nair, Parasuraman Swaminathan, and
  Kavitha Arunachalam.
\newblock Planar printed e-field sensor array for microwave nde of composites.
\newblock In {\em Advances in Non-destructive Evaluation}, pages 219--228.
  Springer, 2021.

\bibitem{Alam2016wearable}
Muhammad~Mahtab Alam, Dhafer~Ben Arbia, and Elyes~Ben Hamida.
\newblock Wearable wireless sensor networks for emergency response in public
  safety networks.
\newblock In {\em Wireless Public Safety Networks 2}, pages 63--94. Elsevier,
  2016.

\bibitem{Hassan2017inkjet}
Arshad Hassan, Shawkat Ali, Gul Hassan, Jinho Bae, and Chong~Hyun Lee.
\newblock Inkjet-printed antenna on thin pet substrate for dual band wi-fi
  communications.
\newblock {\em Microsystem Technologies}, 23(8):3701--3709, 2017.

\bibitem{Jo2014cpw}
Sangjin Jo, Hyunjin Choi, Beomsoo Shin, Sangyeol Oh, and Jaehoon Lee.
\newblock A cpw-fed rectangular ring monopole antenna for wlan applications.
\newblock {\em International Journal of Antennas and Propagation}, 2014, 2014.

\bibitem{Danideh2013cpw}
A~Danideh and RA~Sadeghzadeh.
\newblock Cpw-fed slot antenna for mimo system applications.
\newblock {\em Indian Journal of Science and Technology}, 6(1):3872--3875,
  2013.

\bibitem{Edwards2016foundations}
Terry~C Edwards and Michael~B Steer.
\newblock {\em Foundations for microstrip circuit design}.
\newblock John Wiley \& Sons, 2016.

\bibitem{Chaimool2012cpw}
Sarawuth Chaimool and Prayoot Akkaraekthalin.
\newblock Cpw-fed antennas for wifi and wimax.
\newblock {\em Advanced Transmission Techniques in WiMAX}, page~19, 2012.

\bibitem{Abutarboush2015inkjet}
Hattan~F Abutarboush, Muhammad~Fahad Farooqui, and Atif Shamim.
\newblock Inkjet-printed wideband antenna on resin-coated paper substrate for
  curved wireless devices.
\newblock {\em IEEE Antennas and Wireless Propagation Letters}, 15:20--23,
  2015.

\bibitem{Derby2010inkjet}
Brian Derby.
\newblock Inkjet printing of functional and structural materials: fluid
  property requirements, feature stability, and resolution.
\newblock {\em Annual Review of Materials Research}, 40:395--414, 2010.

\bibitem{Sim2017rf}
Sung-min Sim, Yeonsu Lee, Hye-Lim Kang, Kwon-Yong Shin, Sang-Ho Lee, and
  Jung-Mu Kim.
\newblock Rf performance of ink-jet printed microstrip lines on rigid and
  flexible substrates.
\newblock {\em Microelectronic Engineering}, 168:82--88, 2017.

\bibitem{Chakraborty2019templated}
Anirban Chakraborty, Nitheesh~M Nair, Anjali Adekar, and P~Swaminathan.
\newblock Templated electroless nickel deposition for patterning applications.
\newblock {\em Surface and Coatings Technology}, 370:106--112, 2019.

\end{thebibliography}

%This defines the bibliographies style. Search online for a list of available styles.
%\bibliographystyle{unsrt}
\bibliographystyle{hunsrt}
\end{document}